\documentclass[preprint, superscriptaddress, showpacs,preprintnumbers,amsmath,amssymb,prb]{revtex4}
\usepackage{graphicx}

\begin{document}

\thispagestyle{empty}

\title{Measurement of the gradient of the Casimir force between
a nonmagnetic sphere and a magnetic plate}

\author{ A.~A.~Banishev,${}^1$ C.-C.~Chang,${}^1$
G.~L.~Klimchitskaya,${}^2$ V.~M.~Mostepanenko,${}^3$
and U.~Mohideen${}^1$}
\affiliation{${}^1$Department of Physics and
Astronomy, University of California, Riverside, California 92521,
USA \\
${}^2$North-West Open Technical University,
Polustrovsky Avenue 59,
St.Petersburg, 195597, Russia \\
${}^3$Central Astronomical Observatory
of the Russian Academy of Sciences,
Pulkovskoye chaussee 65/1,
St.Petersburg, 196140, Russia
}

\begin{abstract}
We measured the gradient of the Casimir force between an Au sphere
and a plate made of ferromagnetic metal (Ni). It is demonstrated
that the magnetic properties influence the force magnitude.
This opens prospective opportunities for the control of the
Casimir force in nanotechnology and for obtaining Casimir
repulsion by using ferromagnetic dielectrics.
\end{abstract}

\maketitle

\section{Introduction}

In the last few years the Casimir effect \cite{1} has attracted
much experimental and theoretical attention as a
fluctuation-induced quantum phenomenon with diverse applications
 ranging from nanotechnology to elementary particle physics,
gravitation and cosmology.\cite{2,3}
Many experiments on measuring the Casimir force in
configurations with material boundaries made of nonmagnetic
metals, semiconductors and dielectrics have been performed
(see reviews \cite{4,5,6}).
On the theoretical side, the Lifshitz theory \cite{7}
of the van der Waals and Casimir forces between plane-parallel
plates described by the frequency-dependent dielectric
permittivity $\varepsilon(\omega)$ was generalized for the
boundary surfaces of arbitrary shape \cite{8,9,10}.
Nevertheless, application of this theory to materials
with free charge carriers is under much
discussion.\cite{3,4,6,22,23,24,25,10a,10b,10L,10c,10d,10e,11,12,12a,12b}

The Lifshitz theory has long been generalized \cite{13}
for the case of
magnetic materials described by the frequency-dependent
magnetic permeability $\mu(\omega)$.
The Casimir force between magnetodielectric plates was
studied theoretically by many authors (see,
e.g., \cite{14,15,16,17,18,Rosa,19,19a}), especially in connection
with the possibility of Casimir repulsion, which has
promise for nanotechnology.
A large contribution of magnetic properties to the Casimir
force (including the Casimir repulsion through a vacuum gap)
was predicted \cite{18} for constant $\varepsilon$ and $\mu$.
For real magnetic metals described by the Drude model at
nonzero temperature the effect of repulsion was not
confirmed.\cite{15,19} In Refs.~\cite{15,20} it was shown
that for real magnetic materials at room temperature the
influence of magnetic properties on the Casimir force can
occur solely through the zero-frequency contribution to
the Lifshitz formula (the reason is that $\mu(i\xi)$ drops
to unity at $\xi$ several orders of magnitude lower than the
first Matsubara frequency).
It was also predicted \cite{Rosa,20}
that in the interaction of a nonmagnetic metal with a
ferromagnetic dielectric the Casimir repulsion is possible
if the relaxation properties of free electrons in a metal
and the dc conductivity of the dielectric are omitted.
Keeping in mind that some measurements
\cite{10a,10b,10L,10e,11,12,12a,12b}
have raised questions on whether the
relaxation properties of free charge carriers and dc conductivity
influence the Casimir force, there is a question
pointed out in Ref.~\cite{20}
if the magnetic properties of real materials will influence
the Casimir force.

This paper starts the experimental investigation of the
Casimir effect in the presence of magnetic boundaries.
Using the dynamic atomic force microscope (AFM), operated
in a frequency shift technique, we have measured the
gradient of the Casimir force between an Au-coated
sphere and a plate covered with the ferromagnetic metal Ni.
The experimental data were compared with predictions of
the Lifshitz theory extrapolated to zero frequency both
including \cite{22,23} and omitting \cite{24,25} the
relaxation properties of free electrons (i.e., using the
Drude and plasma model approaches discussed in the
literature.\cite{3,4}) It was found that the measurement
data are in excellent agreement with the predictions of the
plasma model approach taking the magnetic properties into
account and with the Drude model approach (which is not
sensitive to magnetic properties for a nonmagnetic metal
interacting with the magnetic one). Predictions of both these
approaches are shown to be very close within the range of
experimental separations $220\,\mbox{nm}\leq a\leq 500\,$nm.
If to combine the obtained results with an
exclusion of the Drude model approach for
nonmagnetic metals found in several experiments \cite{10a,10b},
one can consider our experiment as a
confirmation of the influence of magnetic properties on the
Casimir force. This opens outstanding opportunities for the
tuning of the Casimir force in nanotechnology by means of
depositing ferromagnetic films on movable parts of
microdevices and even obtaining repulsive Casimir forces
(other means are considered in Refs.\cite{4,5,28,P}).

The paper is organized as follows. In Sec.~II we describe the
measurement scheme and the experimental setup using a dynamic
AFM. Section III is devoted to the calibration procedures and
measurement results. Section IV contains comparison between
experiment and theory. In Sec.~V the reader will find our
conclusions and discussion.

\section{Measurement scheme and experimental setup}

We have applied different voltages $V_i$ to the plate and
measured the gradient of the total force, electric plus
Casimir,
\begin{equation}
F_{\rm tot}(a)=F_{\rm el}(a)+F(a),
\label{eq0}
\end{equation}
\noindent
between an
Au-coated sphere of radius $R$ and Ni-coated plate using
dynamic force microscopy, more specifically,
frequency modulation AFM.\cite{26}
It was originally conceived as a technique to explore
short-range forces. The sensing element of this technique is
the change of resonant frequency of a periodically driven
cantilever, where the change is proportional to the gradient
of the force. It is always assumed that the amplitude of the
oscillation of the cantilever is small. In this technique
the driving frequency is kept near the resonance frequency
of the cantilever to obtain the highest signal to noise.
Note that previous experiments on measuring the Casimir
force by means of dynamic AFM used the phase shift \cite{27} and
the amplitude shift \cite{28} techniques. These have not
been as precise as the frequency modulation technique.
For small oscillations of the cantilever, the change in
the frequency $\Delta\omega=\omega_r-\omega_0$, where
$\omega_r$ is the resonance frequency in the presence of
external force $F_{\rm tot}$ and $\omega_0$ is the
natural resonance frequency, is given by
\begin{equation}
\Delta\omega=-\frac{\omega_0}{2k}\left(
\frac{\partial F_{\rm el}}{\partial a}+
\frac{\partial F}{\partial a}\right)=
-\beta(V_i-V_0)^2-C\frac{\partial F}{\partial a}.
\label{eq1}
\end{equation}
\noindent
Here, $C\equiv C(k,\omega_0)=\omega_0/(2k)$,
$\beta\equiv\beta(a,z_0,C,R)=C\partial X(a,R)/\partial a$,
$k$ is the spring constant of the cantilever, absolute
separations $a=z_{\rm piezo}+z_0$, where $z_{\rm piezo}$ is the
plate movement due to the piezoelectric actuator and $z_0$
is the point of the closest approach between the Au sphere and
Ni plate, and $X(a,R)$ is a known function.\cite{3,4,10a,12}
The residual potential difference $V_0$ in (\ref{eq1}) can be
nonzero even for a grounded sphere due to the different work
functions of the sphere and plate materials or contaminants
on the interacting surfaces.

Expression (\ref{eq1}) is the basis of the technique used in
this experimental study to explore the Casimir effect in magnetic
materials.
The electrostatic force gradient and therefore the frequency shift
has a parabolic dependence on the voltage $V_i$ applied to the plate.
{}From Eq.(\ref{eq1}), the minimum in the frequency shift corresponds
to $V_0$. The curvature of the parabola, which includes the spatial
dependence of the electrostatic force and the cantilever parameters,
is related to $\beta$ and depends on $z_0$ and $C$.
Thus these parameters can be extracted from this dependence.
 In order to test for systematic errors in the fitting parameters,
 the fitting is repeated at many different distance ranges.

The instrument was designed for a precision measurement of the Casimir force
gradient. The primary pieces of equipment necessary for frequency
modulation measurements consist of
a microfabricated
cantilever,  the cantilever motion controllers
(piezoelectric transducers), plate motion controller (piezoelectric tube),
optical detection system,
and a phase locked loop (PLL) to measure the frequency shift.
The cantilever was placed inside a high vacuum chamber. All these allowed
us to reach subnanometer separation distance resolution over a distance
range larger than two micrometers.

The sphere was attached to the cantilever. The chosen sphere needs
to have a large $R$ and a smooth clean surface.
To attach it to the cantilever we used a dot of Ag epoxy at the tip of the
free-end of the rectangular cantilever. To achieve high resonance frequencies
and high signal to noise,  hollow glass microspheres (3M Scotchlite) and stiff
rectangular conductive Si cantilevers were used.
The amount of epoxy has to be minimal
to obtain the high frequencies. We used commercial monocrystalline
cantilevers that are n-type and have a specific resistance of 0.01 to
0.05\,$\Omega\,$cm. We chosen them because they have very low amount
of internal stresses, in comparison to other materials like silicon nitride
(SiN${}_3$). This leads to low energy dissipation, resulting in a high quality
factor $Q$. We chosen ones with the smallest spring constant ($\sim 0.03\,$N/m).
The cantilevers were cleaned with high-purity acetone and then rinsed
with distilled and deionized water. After that, the silicon dioxide
(SiO${}_2$) layer on the surface of the cantilever was etched with a solution of
HF for 1\,min. Finally, they were double rinsed in a solution of
deionized water.
The Si cantilevers are conductive, which is necessary for good electrical contact
to the sphere.

The surfaces of the hollow glass spheres are smooth as they are made from liquid
phase. We used the special procedure for cleaning the spheres before attachment
to the cantilever to remove organic contaminants
and debris from the surface.
Details of this procedure are described in Ref.~\cite{10b}.

The next step was to thermally evaporate a uniform coating of
$280\pm 1\,$nm of Au on the sphere and at the top of the
free end of the cantilever. The Au on the top of the cantilever has to be evaporated only
at the tip, about 50\,$\mu$m from the free end.
The radius of the sphere was measured to be $R=64.1\pm 0.1\,\mu$m using a SEM.
The tip of the cantilever and sphere were coated with Au using a specially designed mask.
Coating only the cantilever tip, preserves the large oscillation $Q$-factor leading to
high sensitivity.  We used an oil-free thermal evaporator for the Au coating. This instrument
is equipped with a scroll pump (``Varian", SH-100) and a turbomolecular pump (``Varian", TV301
Navigator) that permits us to coat at a pressure of  $10^{-6}\,$Torr.   To achieve uniformity
the cantilever sphere system was rotated using a motor during evaporation. To obtain a smooth
coating, the coating rate has to be low and done with cooler atoms. The Au coating was done
over 4 hours, at a boat-cantilever distance of about $2.5^{\prime\prime}$ to achieve a uniform
280\,nm thick layer. To obtain smooth coatings, it was found that an equilibrium pressure be
reached after melting of the Au wire, before start of the coating process.

We used a Si plate as substrate for the Ni coating. The plate was first cleaned in acetone
for 15\,min using a sonicator,  then rinsed with deionized water. This was repeated in IPA for another
15\,min and rinsed with deionized water. Finally it was cleaned in ethanol for 15\,min again using a
sonicator and blow dried in nitrogen. To increase adhesion we placed the substrate in a UV
chamber for 30\,min. Next we inserted the substrate in a E-beam evaporator (``Temescal systems'',
 model BJD-1800) for Ni coating. A vacuum of $10^{-6}\,$Torr was used. The rotation of the sample
 was done to provide the uniform Ni coating. To obtain a coating rate of  $\sim 3\,$\AA/s, the
 high voltage supply controlling the electron beam was kept at 10.5\,kV and filament current at
 0.5\,A. The thickness of the Ni coating was measured with an AFM
to be $154\pm 1\,$nm. The roughness of
the Au and Ni coatings was measured using an AFM at the end of the experiment.

The cantilever with the Au-coated sphere was clamped in a specially fabricated holder
containing two piezos.
The first piezo was connected to a closed proportional-integral-derivative (PID)
controller.
The second was connected
to the PLL. The Ni plate was mounted on top of a $3^{\prime\prime}$ segmented piezoelectric tube
 capable of traveling a distance of $2.3\,\mu$m.
 Using a $3^{\prime\prime}$ segmented piezoelectric tube allows us to achieve large separation between
 the substrate with subnanometer spatial resolution.
 Ohmic contacts were made to the Ni plate through
  a 1\,k$\Omega$  resistor. The calibration of the plate piezo was done using the fiber interferometer
and is described in previous work.\cite{29}
 A continuous 0.01\,Hz triangular voltage
signal was applied to the piezoelectric actuator
to change the sphere-plate
distance and avoid piezo drift and creep.
The experiments were performed in a vacuum of $3\times10^{-8}\,$Torr at room temperature.

The main vacuum chamber was a $8^{\prime\prime}$ six-way stainless steel cross. The chamber was evacuated
by a turbo-pump (``Varian Inc.", V-301) followed by an oil-free dry scroll pump (``Leybold Vac.", SC-15D).
The main vacuum chamber was mounted on a $8^{\prime\prime}$ ion pump (``Varian Inc." Diode). The chamber
was separated from the turbo pump by gate valve ($6^{\prime\prime}$ viton o-ring sealed gate with stainless
steel construction) which can be closed to isolate the turbo pump from the chamber.
During data acquisition only the ion pump was used and the turbo pump was valved off and shut off to
reduce the mechanical noise. The chamber was supported on a damped optical table having a large mass
to reduce the mechanical noise.

We detected the cantilever oscillations and Ni plate displacements with two fiber interferometers.
The first interferometer monitored the cantilever oscillation. The second recorded the displacement of the Ni
plate mounted on the AFM piezoelectric transducer.
These custom-built interferometers consist of a laser
diode pigtail, various types of fiber couplers and photodiodes for detecting the interferometric signal.
Since all junctions between the components of the interferometer were spliced, stray interference and
retroreflection to the laser diodes were minimized.
In addition, to minimize laser and wavelength
fluctuations, the temperature and power of the laser diodes were kept constant using feedback circuits.

Below we describe {the first interferometer}, which was used for monitoring the resonant frequency of the
cantilever. The interferometric cavity was formed with the cleaved end of an optical fiber and the top
free end of the microcantilever.  The cantilever chip holder sits on top of two small piezoelectrics,
one for oscillating the cantilever and the second for changing the length of the interferometric cavity.
For constructing the interferometer, we used a 1550\,nm (``Thorlabs Inc.", 1550B-HP) single mode fiber
which has extremely low bending loss and low splice loss. A super luminescent diode (``Covega Co", SLD-1108)
with a wavelength of 1550\,nm and a coherence length of 66\,$\mu$m
served as the light source
for the cantilever frequency measurement
interferometer.  The short coherence length prevents
spurious interferences. In addition, an
optical isolator with FC-APC connectors connected the diode to a 50/50 directional coupler
to reduce undesirable reflections.
A typical fused-tapered bi-conic coupler at 1550\,nm wavelength with return-loss of
--55dB was used as
input. To position the fiber end vertically above and close to the cantilever, an
$xyz$-stage as described in Ref.~\cite{10b} was used.

{The second interferometer}, which was used for detecting the distance moved by the Ni plate travel,
had the same fabrication techniques. The only difference is the fiber coupled laser source (``Thorlabs Inc.",
S1FC635) with a wavelength of 635\,nm.

Now we consider
the interferometer system used for measuring the frequency shift.  The output  light was measured with
InGaAs photodetectors.
For the cantilever interferometer
a low noise photodetector amplifier system was constructed using a
balanced InGaAs
photodiode coupled to an OPA627 low noise operational amplifier.
The output of the interference signal
was fed into a band-pass filter (``SRS Inc.", SR650, with the range set between $1.20-1.85\,$kHz) cascaded
by low (``SRS Inc.", SR965, 260\,Hz) and high (``SRS Inc.", SR965, 1\,kHz) pass filters to cut off unwanted
frequency bands. The high-pass filter helped us to remove the noise in the excitation signal from
frequency modulation
controller and the low-pass filter cleaned the signal to the PID loop.

 For frequency demodulation we
used a PLL. The PLL frequency demodulator system combines a
controller module to maintain the resonant frequency and a detector module to measure the force gradient
induced resonant frequency shift.
A piezoelectric actuator connected to
the output of the PLL  drives
the cantilever at its resonant frequency with a constant amplitude.
For all separations the oscillation amplitude of the
cantilever was fixed at $<10\,$nm.
To control the amplitude, the fluctuation spectrum of the cantilever
was measured with a spectrum analyzer. The latter was calibrated
using the thermal noise oscillation spectrum which has an average
amplitude of $\sim 0.4\,$nm. During experiment the signal on the
spectrum analyzer was about 15 times higher than for the thermal
noise.
The output of the low pass filter was used to form
a closed loop PID controller using a piezoelectric actuator, to maintain constant separation distance
 between the fiber end and the cantilever.  LabView software was used to control and monitor PID function
 and to acquire the data. The program also sets the voltages applied to the Ni plate using a low-noise
voltage supply.

\section{Calibration procedures and measurement results}

The frequency shift of our oscillator due to the
total force (electrostatic and Casimir) between the Au sphere and the Ni plate was measured as a function
of sphere-plate separation distance. For this purpose, the Ni-coated plate was connected to a voltage
supply (33220A, ``Agilent Inc.") operating with $1\,\mu$V resolution. 11 different voltages in the range
from --87.1 to 25.9\,mV were applied to the Ni plate, while the sphere remained grounded.
The plate was moved towards the sphere
starting at the maximum separation,  and the corresponding frequency
shift was recorded at every 0.14\,nm.
This measurement was repeated four times.
Any mechanical drift was subtracted from the separation distance.
To do so we took into account that at $a>2\,\mu$m, the total force between the sphere
and plate is below the instrumental sensitivity. At these separations, the noise is far greater than the
signal and in the absence of systematic errors the signal should average to zero.
However, the drift caused the separation distance to increase by around 1\,nm in 1000\,s, where 100\,s
correspond to time taken to make the one measurement (note that positional precision much better than 1\,nm
was achieved in this experiment).
The drift rate was calculated from the change in position at one frequency
shift signal plotted as a function of time.
This procedure was repeated for 15 different frequency shift signals to calculate
the average drift rate, which was found to be 0.002\,nm/s.
The separation distance in all measurements
was corrected for this drift rate.
Note that the temperature was controlled through excellent
thermal contact to the heat bath, which is the large thermal mass
of the vacuum chamber and optical table. The experiment was
equilibriated for several hours before data were taken.
The drift was experimentally
found to be linear during 40\,min, which is the time scale of
the measurements
(see Ref.\cite{10b} for details where the same setup was used).

After applying the drift correction the residual potential $V_0$ between the sphere and plate was
determined using the following procedure.
For every 1\,nm separation the frequency shift signals  were found
by interpolation.
Then these signals were plotted as a function of the applied voltage $V_i$ at every separation and
the corresponding $V_0$ identified at the position of the parabola
maxima.
The curvature of the parabola $\beta$ was also found. These $V_0$ from all four measurement sets are plotted in
Fig.~1(a)  versus separation distance.
In Fig.~1(b) we show the systematic error of each individual $V_0$,
as determined from the fit, at different separations.
The mean value is $V_0=-34.1\pm 1\,$mV, where the total error is
determined at a 67\% confidence level, and can be observed to be
independent of separation.
This is an indirect confirmation of the fact that the interacting regions of
the surfaces are clean or the adsorbed impurities are randomly distributed with a sub-micrometer length
scales and do not contribute to the total force.\cite{10b}

The next step was to determine the separation distance at closest approach $z_0$ and the coefficient $C$
in Eq.~(\ref{eq1}). As discussed above, these parameters associated with the cantilever can be found from
the dependence of the parabola curvature $\beta$ on distance. The corresponding theoretical expression for
 parabola curvature was fit to $\beta$ as function of the separation distance.
 A least $\chi^2$ procedure was used in the fitting and the best values of $z_0$ and $C$ were obtained.
The fitting procedure was repeated by keeping the start point fixed at the closest separation, while the
end point measured from the closest separation was varied from 750 to 50\,nm.
The values of $z_0$  shown in Fig.~2(a) are seen to be independent of the end
point position indicating the absence of systematic errors resulting from $z_{\rm piezo}$ calibration,
mechanical drift etc.
The systematic errors of each individual $z_0$, as determined from
the fit, vary between 0.32 and 0.45\,nm.
Similarly, the values of the coefficient $C$
shown in Fig.~2(b) were also extracted
by fitting the $\beta$-curve as a function of separation.
The systematic errors of each individual $C$ vary between 0.13
and 0.18\,kHz\,m/N.
The independence of $C$ on separation again indicates
the absence of systematic errors. The mean values obtained were $z_0=217.1\pm 0.4$\,nm and
$C=46.7\pm 0.15\,$kHz\,m/N, respectively (see Ref.\cite{10b} for
details).
After the determination of $z_0$ the absolute sphere-plate separations can be found.
Using $C$ and the measured frequency shifts, the gradients of the Casimir force
were calculated from Eq.~(\ref{eq1}).

We now turn to the determination of the experimental errors.
The random error in the gradient of the Casimir force calculated
from 44 repetitions (4 measurement sets with 11 applied voltages
each) at a 67\% confidence level is shown by the short-dashed line
in Fig.~3. The systematic error is determined by the instrumental
noise including the background noise level, by the errors in
calibration and by the error in the gradient of electrostatic
force subtracted in accordance with Eq.~(\ref{eq1}) to get
$F^{\prime}(a)$. It is shown by the long-dashed line
in Fig.~3.
The total experimental error found at a 67\% confidence level
is shown by the solid line in Fig.~3. The error in separation
$\Delta a=0.4\,$nm coincides with the error in the determination
of $z_0$ (details of error analysis can be found in
Ref.\cite{10b}).

In Fig.~4(a-d) the measured gradients of the Casimir force are
indicated as crosses, where the arms of the crosses are
determined by the total experimental errors. It can be seen
that the total relative experimental error in the measurements
of $F^{\prime}(a)$ at $a=220$, 250, 300, 400, and 500\,nm is
equal to 0.6\%, 0.94\%, 1.8\%, 5.4\%, and 12.8\%, respectively.

\section{Comparison between experiment and theory}

Now we compare the experimental data for the gradient of the
Casimir force in the configuration containing nonmagnetic (Au)
and magnetic (Ni) metals with the  predictions of the Lifshitz
theory. Computations were performed using the
following Lifshitz-type
formula \cite{20} for $F^{\prime}(a)$ obtained in the
sphere-plate geometry with the help of the proximity force
approximation (using the exact theory, it was recently
shown \cite{30,31} that the error in this case is less
than $a/R$, i.e., less than 0.3\% at
the shortest separation):
\begin{eqnarray}
&&
F^{\prime}(a,T)=2k_BTR
\sum_{l=0}^{\infty}{\vphantom{\sum}}^{\prime}
\int_{0}^{\infty}q_lk_{\bot}dk_{\bot}
\label{eq1a} \\
&&~
\times
\left[
\frac{r_{\rm TM}^{\rm (Au)}r_{\rm TM}^{\rm (Ni)}}{e^{2aq_l}-
r_{\rm TM}^{\rm (Au)}r_{\rm TM}^{\rm (Ni)}}+
\frac{r_{\rm TE}^{\rm (Au)}r_{\rm TE}^{\rm (Ni)}}{e^{2aq_l}-
r_{\rm TE}^{\rm (Au)}r_{\rm TE}^{\rm (Ni)}}
\right].
\nonumber
\end{eqnarray}
\noindent
Here, $k_B$ is the Boltzmann constant, $T=300\,$K is
the temperature at the laboratory,
$q_l^2=k_{\bot}^2+\xi_l^2/c^2$, and
$\xi_l=2\pi k_BTl/\hbar$ with $l=0,\,1,\,2,\,\ldots$ are
the Matsubara frequencies.
The prime near the summation sign multiplies the term with
$l=0$ by 1/2. The reflection coefficients for transverse
magnetic (TM) and transverse electric (TE) polarizations
of the electromagnetic field on a Ni plate are given by
\begin{eqnarray}
&&
r_{\rm TM}^{\rm (Ni)}=\frac{\varepsilon^{\rm (Ni)}(i\xi_l)q_l-
k_l^{\rm (Ni)}}{\varepsilon^{\rm (Ni)}(i\xi_l)q_l+k_l^{\rm (Ni)}},
\nonumber \\
&&
r_{\rm TE}^{\rm (Ni)}=\frac{\mu^{\rm (Ni)}(i\xi_l)q_l-
k_l^{\rm (Ni)}}{\mu^{\rm (Ni)}(i\xi_l)q_l+k_l^{\rm (Ni)}},
\label{eq1b}
\end{eqnarray}
\noindent
where
\begin{equation}
k_l^{\rm (Ni)}=\left[ k_{\bot}^2+
\varepsilon^{\rm (Ni)}(i\xi_l)\mu^{\rm (Ni)}(i\xi_l)
\frac{\xi_l^2}{c^2}\right]^{1/2}.
\label{eq1c}
\end{equation}
\noindent
The reflection coefficients on Au are obtained from
Eqs.~(\ref{eq1b}) and (\ref{eq1c}) by replacing the
dielectric permittivity of nickel $\varepsilon^{\rm (Ni)}$
with the dielectric permittivity of gold $\varepsilon^{\rm (Au)}$
and the magnetic permeability of nickel $\mu^{\rm (Ni)}$
with unity.

The dielectric permittivity of Au along the imaginary
frequency axis was obtained by means of the Kramers-Kronig
relation from the tabulated optical data \cite{32}.
The latter were extrapolated to zero frequency by means of
the Drude model with the plasma frequency
$\omega_{p,{\rm Au}}=9.0\,$eV and the relaxation parameter
$\gamma_{\rm Au}=0.035\,$eV (the Drude model approach) or
with the relaxation properties of conduction electrons
omitted and using the extrapolation to zero frequency by means
of the simple plasma model with the same $\omega_{p,{\rm Au}}$
(the plasma model approach). The dielectric properties of Ni
were described in two similar ways, but with
$\omega_{p,{\rm Ni}}=4.89\,$eV and
$\gamma_{\rm Ni}=0.0436\,$eV.\cite{32,33}
The magnetic properties of the Ni film were described by the static
magnetic permeability $\mu^{\rm (Ni)}(0)=110$ (our sample did not possess
a spontaneous magnetization due to the sufficiently thick Ni
coating used and weak environmental magnetic fields in our
experimental setup). As was mentioned above, at
room temperature only the contribution from zero Matsubara
frequency may be influenced by the magnetic properties of
a material.\cite{20} At separations above 220\,nm
considered here the influence of surface roughness onto the
gradient of the Casimir force can be taken into account
within the multiplicative approach. {}From the AFM scans
the r.m.s. roughness of the sphere and plate surfaces
was measured to be $\delta_s=2.0\,$nm
and $\delta_p=0.49\,$nm, respectively, leading to a multiple
equal to 1.0009 at $a=220\,$nm. Thus, the contribution of
surface roughness in this experiment is negligibly small.

The computational results for the gradient of the Casimir
force obtained using the plasma model approach with the
magnetic properties of Ni included or omitted [i.e.,
putting $\mu^{\rm (Ni)}(0)=1$] are shown as the solid and dashed
bands in Fig.~4(a-d). The widths of the bands indicate
the theoretical error. The difference between the two
bands is explained by the fact that the reflection
coefficient with TE polarization on a Ni plate
described by the plasma model at zero Matsubara frequency
depends on $\mu^{\rm (Ni)}(0)$:
\begin{equation}
r_{{\rm TE},p}^{(\rm Ni)}(0,k_{\bot})=
\frac{\mu^{\rm (Ni)}(0)ck_{\bot}-\left[c^2k_{\bot}^2
+\mu^{\rm (Ni)}(0)\omega_{p,{\rm Ni}}^2\right]^{1/2}}{\mu^{\rm (Ni)}(0)ck_{\bot}
+\left[c^2k_{\bot}^2+\mu^{\rm (Ni)}(0)\omega_{p,{\rm Ni}}^2\right]^{1/2}},
\label{eq2}
\end{equation}
\noindent
where $k_{\bot}$ is the projection of the wave vector on the
plate. At the same time
$r_{{\rm TE},p}^{(\rm Au)}(0,k_{\bot})\neq 0$.
As can be seen in Fig.~4, the plasma model approach with
magnetic properties included is in excellent agreement with
the data over the entire measurement range.
The plasma model approach with the magnetic properties omitted
is excluded by the data at a 67\% confidence level over the
interaction range from 220 to 420\,nm.

The Drude model approach is not sensitive to the presence of
magnetic properties in this experiment because
$r_{{\rm TE},D}^{(\rm Au)}(0,k_{\bot})= 0$ and, thus,
the magnetic properties do not contribute to $F^{\prime}(a,T)$
regardless of the value of
\begin{equation}
r_{{\rm TE},D}^{(\rm Ni)}(0,k_{\bot})=
\frac{\mu^{\rm (Ni)}(0)-1}{\mu^{\rm (Ni)}(0)+1}
\label{eq2a}
\end{equation}
\noindent
(we note that these coefficients enter the Lifshitz formula
(\ref{eq1a})
as a product). By coincidence, over the separation region from
220 to 500\,nm the predictions of the Drude model approach
almost coincide with the solid line in Fig.~4 (the magnitudes
of relative differences at separations of 220, 300, 400, and
500\,nm are only 0.5\%, 0.2\%, 0.4\%, and 1.2\%, respectively).
Thus, for a nonmagnetic metal interacting with a magnetic one
at short separations, the predictions of both approaches are
much closer than for two Au test bodies where the respective
differences were resolved experimentally.\cite{10a,10b}
Note that at large separations the relative differences between
the Drude model approach and the plasma model approach with
magnetic properties included are much larger (31.1\% and
42.8\% at separations 3 and $5\,\mu$m, respectively).

\section{Conclusions and discussion}

In this work we have measured the gradient of the Casimir force
between an Au-coated sphere and a Ni-coated plate using an AFM
operating in the dynamic regime. We have compared the mean
gradients of the Casimir force with theoretical predictions of
the Lifshitz theory with no fitting parameters. In so doing,
both the Drude and the plasma model approaches to the description
of dielectric properties of metals have been used.
It was found that the experimental data are in excellent agreement
with the plasma model approach with magnetic properties included
and exclude the same approach with the magnetic properties
omitted at a 67\% confidence level. Our experimental data are
also consistent with the Drude model approach which is not
sensitive to the presence of magnetic properties in the
configuration of an Au-coated sphere and Ni-coated plate at
separations considered.

It is pertinent to note that previous
experiments with two Au surfaces performed using a
micromachined oscillator\cite{10a} and a dynamic
AFM\cite{10b} (also used in this experiment) cannot
be reconciled with the Drude model approach.
The experiment using a torsion pendulum has been
claimed\cite{10L} to be in favor of the Drude model approach,
but has been shown\cite{10c,10d} to be not informative at short separations
and in better agreement with the plasma model at large
separations  above $3\,\mu$m.
As a result, we have many reasons to conclude that our work
pioneers measurement of the
influence of magnetic properties of ferromagnets on the
Casimir force. This opens opportunities for the
control of the Casimir forces in nanotechnology and even for
realization of the Casimir repulsion through the vacuum gap
using ferromagnetic dielectrics.

\section*{Acknowledgments}
This work was supported by the DOE Grant
No.~DEF010204ER46131 (equipment, G.L.K., V.M.M., U.M.),
NSF Grant
No.~PHY0970161 (G.L.K., V.M.M., U.M.),
and DARPA Grant under Contract
No.~S-000354 (A.B., U.M.).

\begin{figure*}[h]
\vspace*{-4.0cm}
\centerline{\hspace*{1cm}
\includegraphics{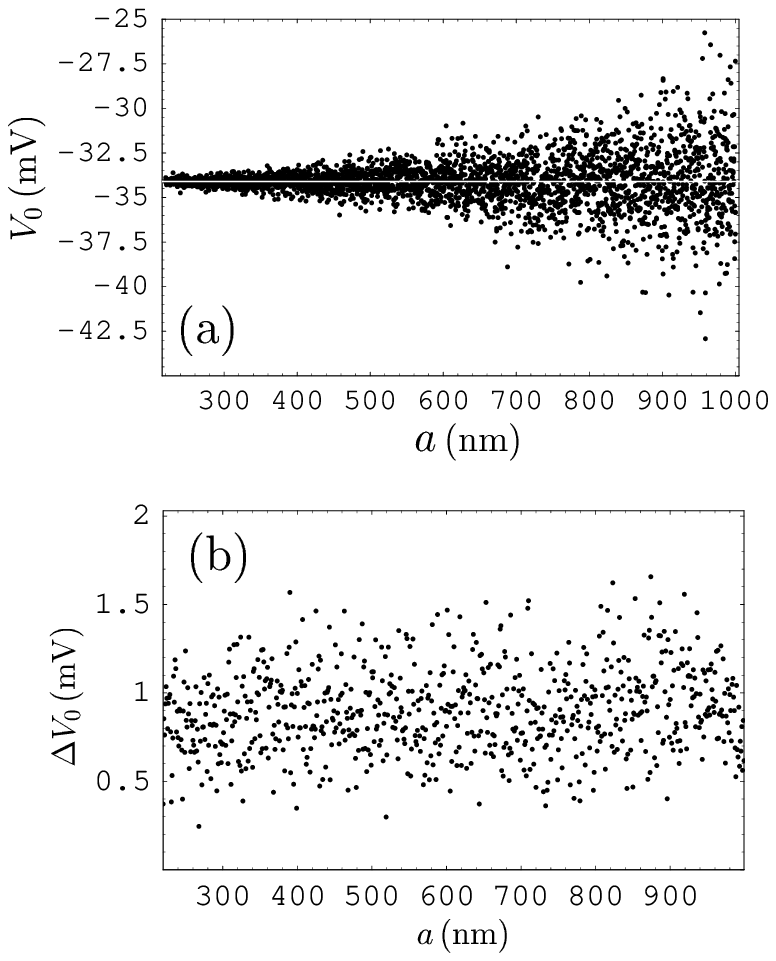}
}
\vspace*{-11.5cm}
\caption{(a) The values of the residual potential $V_0$
found at each separation distance are shown as dots.
The mean value of $V_0$ is shown by the white line.
(b) The systematic error of each individual $V_0$, as determined
from the fit, versus separation.
}
\end{figure*}
\begin{figure*}[h]
\vspace*{-8.cm}
\centerline{\hspace*{1cm}
\includegraphics{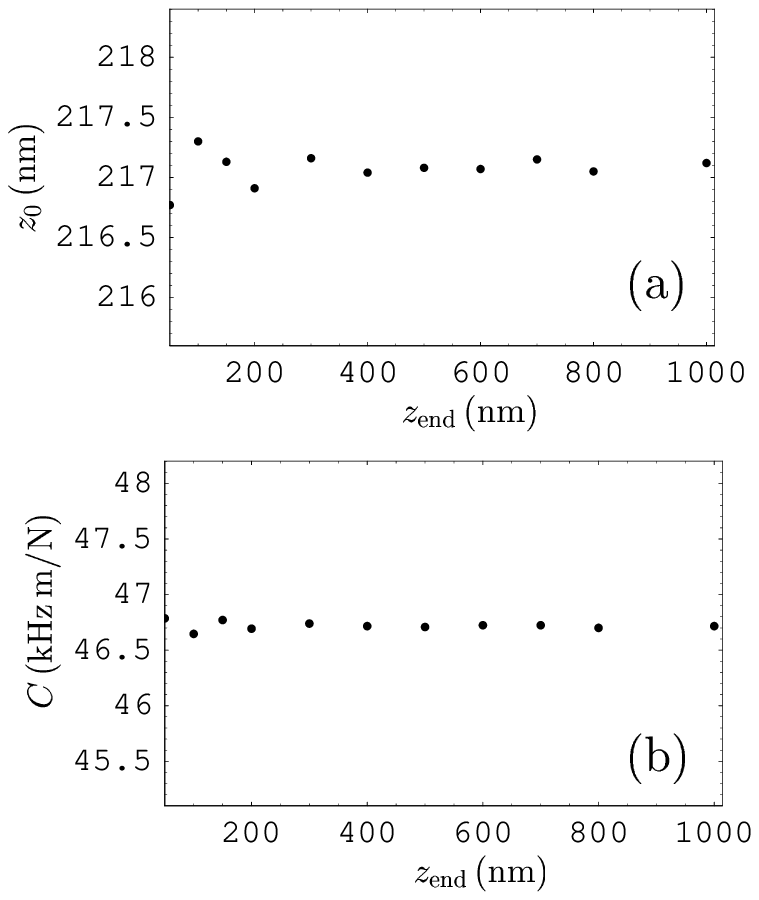}
}
\vspace*{-11.5cm}
\caption{The dependences of (a) the closest sphere-plate separation and
(b) the coefficient $C$ in Eq.~(\ref{eq1}) on the end point of the fit.
}
\end{figure*}
\begin{figure*}[h]
\vspace*{-12.0cm}
\centerline{\hspace*{1cm}
\includegraphics{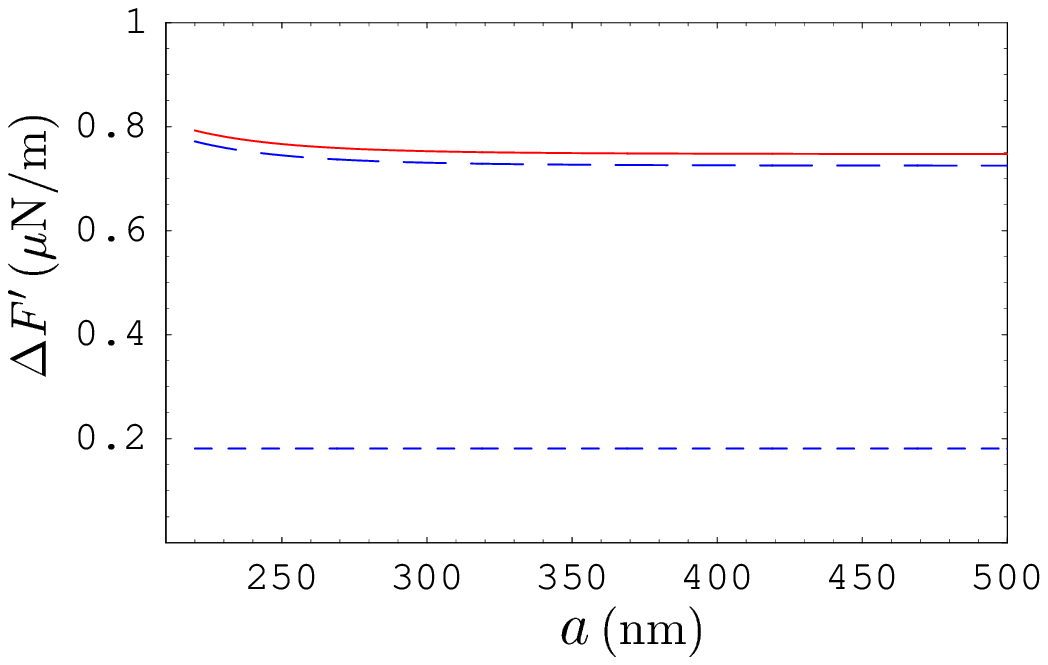}
}
\vspace*{-6.cm}
\caption{(Color online)
The random, systematic and total experimental errors in
the gradient of the Casimir force determined at a 67\%
confidence level are shown by the short-dashed,
long-dashed and solid lines, respectively, as functions
of separation.
}
\end{figure*}
\begin{figure*}[h]
\vspace*{-7.0cm}
\centerline{\hspace*{0cm}
\includegraphics{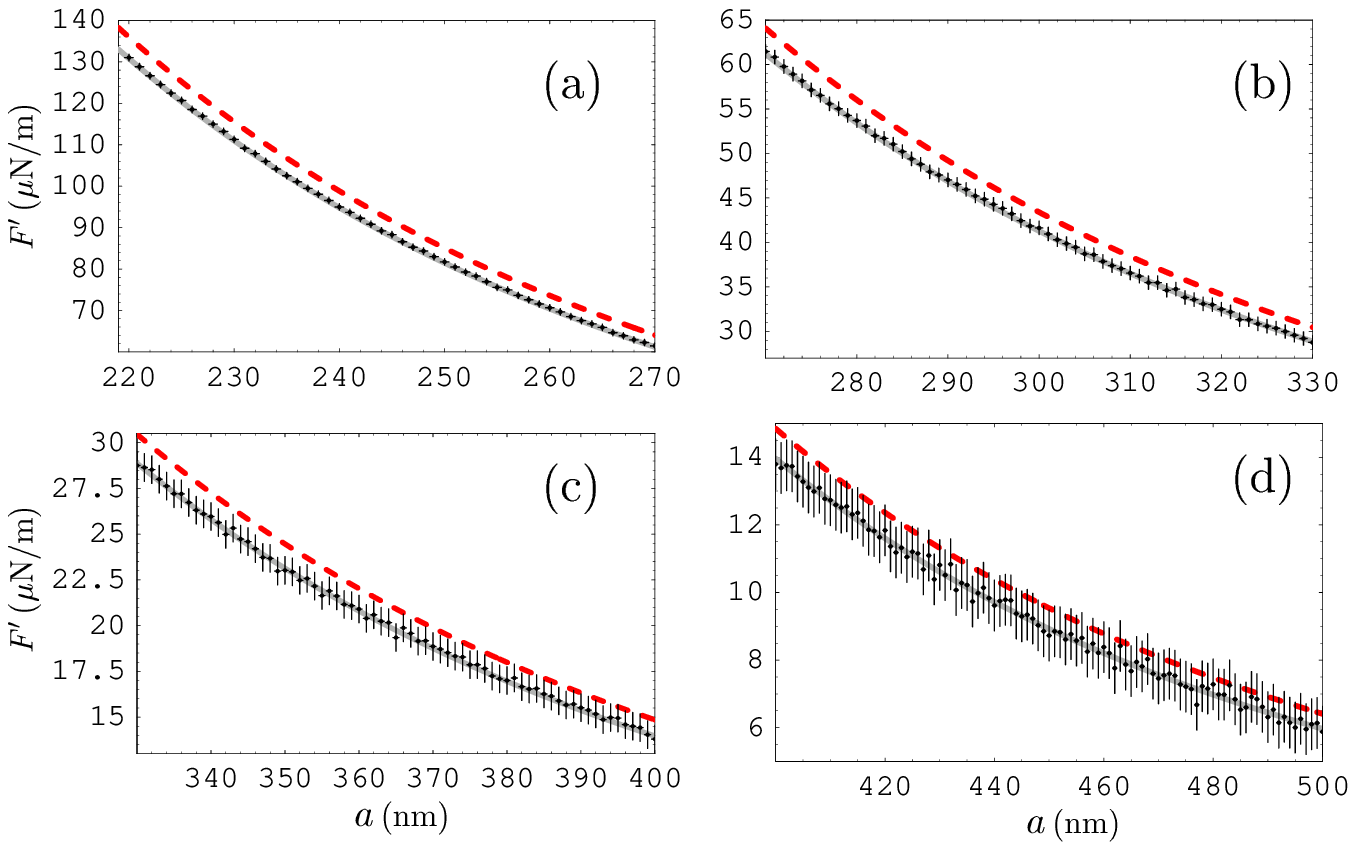}
}
\vspace*{-10.cm}
\caption{(Color online)
Comparison between the experimental data for
the gradient of the Casimir force (crosses) and
theory (solid and dashed bands computed using the
plasma model approach with included and omitted
magnetic properties of Ni plate, respectively).
}
\end{figure*}
\end{document}